\input ./style/arxiv-general.cfg
\documentclass[aoas,MSNbibl,nameyear,dvips]{arximspdf}
\makeatletter
   \@ifpackageloaded{graphicx}{}{\usepackage{graphicx}}
\makeatother

\doi{10.1214/15-AOAS818}
\volume{9}
\issue{2}
\pubyear{2015}
\firstpage{687}
\lastpage{713}
\docsubty{FLA}

\makeatletter
\newcommand{\rrvert}{\vert}
\newcommand{\llvert}{\vert}
\renewcommand{\mid}{|}
\newcommand{\bX}{\mathbf{X}}
\newcommand{\bY}{\mathbf{Y}}
\newcommand{\bw}{\mathbf{w}}
\newcommand{\bbeta}{\bolds{\beta}}
\newcommand{\bgamma}{\bolds{\gamma}}
\newcommand{\bep}{\bolds{\varepsilon}}
\newcommand{\bB}{\mathbf{B}}
\newcommand{\normal}{\mathrm{N}}
\newcommand{\Var}{\mathbb{V}\mathrm{ar}}
\newcommand{\one}{\mathbf{1}}
\newcommand{\bE}{\mathbb E} 
\newcommand{\bV}{\mathbb V}
\makeatother

\begin{document}
\begin{frontmatter}

\title{Spatial Bayesian variable selection and grouping for high-dimensional scalar-on-image regression}
\runtitle{Spatial Bayesian variable selection and grouping}

\begin{aug}
\author[A]{\fnms{Fan}~\snm{Li}\corref{}\thanksref{M1,T11,T1}\ead[label=e1]{fli@stat.duke.edu}},
\author[B]{\fnms{Tingting}~\snm{Zhang}\thanksref{M2,T22,T1}\ead[label=e2]{tz3b@virginia.edu}},
\author[A]{\fnms{Quanli}~\snm{Wang}\thanksref{M1}\ead[label=e3]{quanli@stat.duke.edu}}, 
\author[C]{\fnms{Marlen~Z.}~\snm{Gonzalez}\thanksref{M2}\ead[label=e4]{mzg7uv@virginia.edu}},
\author[C]{\fnms{Erin L.}~\snm{Maresh}\thanksref{M2}\ead[label=e5]{elm2cg@virginia.edu}}
\and
\author[C]{\fnms{James A.}~\snm{Coan}\thanksref{T33,M2}\ead[label=e6]{jcoan@virginia.edu}}
\runauthor{F. Li et~al.}
\affiliation{Duke University\thanksmark{M1} and University of Virginia\thanksmark{M2}}
%
\address[A]{F. Li\\
Q. Wang\\
Department of Statistical Science\\
Duke University\\
Durham,  North Carolina 27708-0251\\
USA\\
\printead{e1}\\
\phantom{E-mail: }\printead*{e3}}
\address[B]{T. Zhang\\
Department of Statistics\\
University of Virginia\\
Charlottesville, Virginia 22904\\
USA\\
\printead{e2}}
\address[C]{M.~Z. Gonzalez\\
E.~L. Maresh\\
J.~A. Coan\\
Department of Psychology\\
University of Virginia\\
Charlottesville, Virginia 22904\\
USA\\
\printead{e4}\\
\phantom{E-mail: }\printead*{e5}\\
\phantom{E-mail: }\printead*{e6}}
\end{aug}
\thankstext{T11}{Supported in part by the U.S. NSF-DMS Grant 1208983.}
\thankstext{T22}{Supported in part by the U.S. NSF-DMS Grants 1209118
and 1120756.}
\thankstext{T33}{Supported in part by the National Institute of Mental
Health (NIMH) Grant R01MH080725.}
\thankstext{T1}{Equally contributing authors.}

%
\received{\smonth{10} \syear{2014}}
%
\revised{\smonth{2} \syear{2015}}

%
\begin{abstract}
Multi-subject functional magnetic resonance imaging (fMRI) data has been
increasingly used to study the population-wide relationship between human
brain activity and individual biological or behavioral traits. A common method
is to regress the scalar individual response on imaging predictors,
known as a
scalar-on-image (SI) regression. Analysis and computation of such
massive and
noisy data with complex spatio-temporal correlation structure is challenging.
In this article, motivated by a psychological study on human affective feelings
using fMRI, we propose a joint Ising and Dirichlet Process (Ising-DP)
prior within
the framework of Bayesian stochastic search variable selection for
selecting brain
voxels in high-dimensional SI regressions. The Ising component of the
prior makes
use of the spatial information between voxels, and the DP component
groups the
coefficients of the large number of voxels to a small set of values and
thus greatly
reduces the posterior computational burden. To address the phase
transition phenomenon
of the Ising prior, we propose a new analytic approach to derive bounds
for the hyperparameters,
illustrated on 2- and 3-dimensional lattices. The proposed method is
compared with
several alternative methods via simulations, and is applied to the fMRI
data collected from the KLIFF hand-holding experiment.
\end{abstract}

%
\end{frontmatter}

\section{Introduction}
Positive social contact is known to enhance human health and
well-being, possibly because it helps to regulate humans' emotional
reactivity when facing negative stressors in daily life [\citet{CoaSchDav06};
\citet{CoaBecAll13}; \citeauthor{Coa10} (\citeyear{Coa10,Coa11})]. Conventional studies on social
contact primarily focus on its aggregated effect on an entire
population. With the common belief that human behavior is controlled by
individual mental decisions, which is affected by the immediate
environment, it is desirable to investigate emotion regulation activity
of the individual brain under different social interaction conditions.
Toward this aim, the KLIFF hand-holding psychological experiment [\citet{CoaSchDav06}]
was conducted. In this experiment, 104 pairs---each pair
consisting of a male and a female---of mentally and physically healthy
young adults in various close relationships including friends and
married couples were recruited from a larger representative
longitudinal community sample [\citet{Alletal07}]. One participant of
each pair was threatened with mild electric shock during a functional
magnetic resonance imaging (fMRI) session while either holding a hand
of a friend, holding a hand of a stranger or holding no hand at all, in
three separate sessions, which represent three different types of
social interactions---positive and supportive social interaction with
friends, general social interaction with strangers and no social
interaction, respectively. At the end of each session, the subjects
were asked to rate their feelings of arousal and valence [\citet{Rus80}; \citet{Lanetal93}] experienced during the experiment.
Arousal and valence are the two dimensions in the framework of emotion
fields, representing the extent of excitement and pleasure experienced,
respectively [see \citet{BraLan94} for more detailed explanation].

To investigate which areas in the brain are predictive of individual's
affective feelings in the KLIFF study, we can construct a regression
model using subjects' emotion (arousal and valence) measurements as the
response, and summaries of the fMRI images in the regions of interests
(ROIs) as predictors. This type of regression is often referred to as
scalar-on-image (SI) regressions in the literature [\citet{Reietal11}; \citet{Huaetal13};
\citet{GolHuaCra14}]. SI regressions
with predictors from other imaging modalities, such as diffusion tensor
imaging (DTI), have also been used in medical and scientific studies
[e.g., \citet{Reietal14}].

The SI regression model in the KLIFF study has several unique
characteristics due to the features of fMRI data. First, the sample
size is much smaller than the number of predictors, that is, the number
of brain voxels (3D cubic volumes in the brain) in the ROIs, which is
over 6000 in the KLIFF study. This is known as the ``large $p$, small
$n$'' paradigm [\citet{Wes03}]. Second, there is rich spatial information
between the predictors. Third, neighboring predictors are highly
correlated and often have similar but weak effects on the response.
Finally, as each voxel accounts for only a tiny area in the brain, it
is very likely that the number of significant voxels is much larger
than the sample size. The last two characteristics imply that even with
all the true voxels being correctly selected, standard regression
methods may still not be applicable due to multicolinearity. It is
therefore desirable to impose a certain degree of shrinkage or grouping
of the regression coefficients so that predictors with similar values
can be grouped together, and thus the effective number of selected
predictors is smaller than the sample size. Motivated by these
considerations, in this article, we propose a Bayesian SI regression
model that achieves simultaneous grouping and spatial selection of
voxels that are predictive of individual responses. The key to our
proposal is to define a joint Ising and Dirichlet Process (Ising-DP)
prior for the regression parameters, within the framework of Bayesian
stochastic search variable selection [SSVS; \citeauthor{GeoMcC93} (\citeyear{GeoMcC93,GeoMcC97})].
The Ising component of the prior utilizes the spatial
information between voxels to smooth the selection indicators of
neighboring voxels, and the DP component groups the coefficients of
voxels with similar effects to improve prediction power and also reduce
the posterior computational burden. This method has scientific,
statistical and computational advantages over several existing
alternative priors.

Bayesian inference has become increasingly popular in fMRI data
analysis due to several attractive properties: first, the posterior
inference offers direct probabilistic interpretation of the estimates;
second, it eschews the multiple-comparison problem faced by classical
inference; third, incorporating prior information is straightforward
within the Bayesian framework. In particular, Markov Random Fields
priors, such as the Ising prior and the Potts prior, have been widely
used to account for the spatial information between voxels [e.g.,
\citet{GosAueFah01}; \citet{Wooetal04}; \citet{PenTruFri05};
\citet{Bow07}; \citet{Bowetal08}; \citet{DerBowKil10};
\citet{Geetal14}] and for meta-analysis [e.g., \citet{Kanetal11};
\citet{YueLinLoh12}]. \citet{Johetal13} used a joint Dirichlet Process mixture
and Potts prior to achieve simultaneous clustering and selection.
Within the SSVS framework, \citet{Smietal03} and \citet{SmiFah07} used the Ising prior in the context of massive univariate
general linear models [GLM, \citet{Frietal95}] for identifying
brain regions activated by a stimulus. It is important to stress that
the setting in Smith and colleagues is fundamentally different from the
SI regression in this paper: the former only involves fMRI time series,
without individual scalar outcome, and it deals with selecting and
smoothing the coefficients from $p$ one-dimensional regressions (one
for each voxel), a setting broadly belonging to multiple testing;
whereas our paper deals with variable selection from one
$p$-dimensional regression, a much more challenging task.

Within the SSVS but outside the fMRI literature, there is a stream of
recent work on using the Ising prior to incorporate existing structure
information between variables under the ``large $p$, small $n$''
paradigm [e.g., \citet{LiZha10}; \citet{Stietal11}; \citet{VanSti11}]. Moreover, simultaneous selection and clustering in
multiple regression was discussed in \citet{TadShaVan05},
\citet{KimTadVan06} and \citet{DunHerEng08}, but none of those incorporated
existing structure between covariates. Another important but
under-investigated issue is phase transition in the Ising model [for a
review, see \citet{Sta87}], which, in the context of variable
selection, leads to a drastic change (from nearly none to nearly all)
in the number of variables selected given an infinitesimal change in
the hyperparameters. And the difficulty and sensitivity in
hyperparameter selection increases substantially as the degree of the
underlying graph increases. Since the fMRI voxels naturally overlay a
3-dimensional lattice, it is crucial to select hyperparameters that
avoid phase transition for valid inference and feasible computation.
However, despite being intensively explored in statistical physics,
phase transition and the consequent issue of hyperparameter selection
has received relatively little attention in the literature of variable
selection. \citet{LiZha10} derived a ballpark estimate of the phase
transition boundary for the Ising prior using mean field theory. But
their derivation is solely based on the prior distribution and does not
take into account the data or any prior knowledge of the predictors,
and thus the resulting range of possible hyperparameters is often very
wide. In this article we develop a new analytic approach to derive a
tighter boundary of the hyperparameters based on the data and the
posterior distribution, and illustrate it on 2- and 3-dimensional lattices.

The rest of the article is organized as follows. Section~\ref{secmodel} introduces the new Bayesian model and Section~\ref{secparselection} develops an analytic approach to hyperparameter
selection. Posterior computation of the model is discussed in Section~\ref{secposterior}. Section~\ref{secsimulation} compares the
proposed methods with several existing methods through simulations. In
Section~\ref{secapplication} we apply the proposed method to the
KLIFF study to investigate the social regulation of human emotion.
Section~\ref{secconclusion} concludes.

\section{The model} \label{secmodel}
We formulate the problem via a standard multiple regression
\begin{equation}
\label{reg} \bY=\bX\bolds{\eta} +\bep,
\end{equation}
where $\bY$ is the $n\times1$ variable response, for example, the
scalar arousal or valence measurement in the KLIFF study; $\bX=(\bX
_1,\dots,\bX_p)$ is the $n\times p$ ($p\gg n$) matrix of spatially
correlated neuroimaging covariates, for example, the magnitudes of the
estimated hemodynamic response function (HRF) of the voxels in the two
ROIs in the study; and $\bep$ is the error term with $\bep\sim\mbox
{N}(0, \sigma^2I_n)$. To focus on the main message, we do not consider
design variables, such as age and sex, which can be easily added to the
regression.

To select the voxels that are predictive of the response, we adopt the
Bayesian SSVS approach that assumes the ``spike-and-slab'' type of
mixture prior for the regression coefficients [\citet{MitBea88};
\citeauthor{GeoMcC93} (\citeyear{GeoMcC93,GeoMcC97}); \citet{SmiKoh96}].
Specifically, we define a latent indicator $\gamma_j \in\{0,1\}$ for
each covariate that indicates whether this covariate is included in the
model (i.e., whether a voxel is significantly predictive of the
response). We let
\[
\label{betaprior2} \eta_j= \gamma_j\cdot
\beta_j\quad\mbox{and}\quad \beta_j\sim G,
\]
where $\beta_j$ represents the regression coefficient of predictor $j$
once it is selected, and $G$ is a prespecified probability
distribution. Given $\gamma_j$ and $G$, $\eta_j$ are independent
following a spike-and-slab prior
\begin{equation}
\label{betaprior} \eta_j\mid (\gamma_j, G) \sim(1-
\gamma_j)\delta_0+ \gamma_j G,
\end{equation}
where $\delta_0$ is a point mass at 0. Our goal is to propose a new
joint Ising and DP (Ising-DP) prior, where an Ising prior is imposed on
$\bgamma=(\gamma_{1},\ldots,\gamma_{p})'$ to incorporate spatial
information between voxels, and, in parallel, a Bayesian nonparametric
DP prior is imposed on $G$ to achieve grouping of the regression
coefficients, as elaborated below.

We represent the spatial structure among the fMRI voxels via a graph.
Let $i\sim j$ denote that $i$ and $j$ are neighboring voxels. Let
$\mathcal{E} = \{(j_1,j_2)\dvtx  1\leq j_1 \sim j_2 \leq p\}$ be the set of
all the neighboring pairs of voxels---the edge set of the underlying
graph. Given $\mathcal{E}$, let $\mathbf{a} = (a_1,\dots,a_p)'$ be\vspace*{1pt} a
vector and $\bB= (b_{j_1,j_2})_{p\times p}$ be a symmetric matrix of
real numbers where $b_{j_1,j_2}=0$ for all $(j_1,j_2) \notin\mathcal
{E}$. To incorporate the prior structural information into the model
building process, we assume an Ising prior distribution for $\bolds
{\gamma}$ [\citet{LiZha10}] as the first component of the proposed prior:
\begin{equation}
\Pr(\bolds{\gamma}) =\exp\bigl\{\mathbf{a}'\bolds{\gamma} + \bolds{
\gamma}'\bB \bgamma-\psi(\mathbf{a},\bB)\bigr\}, \label{isingprior}
\end{equation}
where $\psi(\mathbf{a},\bB)$ is the normalizing constant: $\psi(\mathbf
{a},\bB) = \log \{ \sum_{\bolds{\gamma} \in
\{0,1\}^p} \exp(\mathbf{a}'\bolds{\gamma} + \bolds{\gamma}'\bB\bolds{\gamma
})  \}$. If $\bB=0$, then $\psi(\mathbf{a},\bB)=\sum_{j=1}^p\log
(1+e^{a_j})$, but in general there is no closed form for $\psi$. The
Ising model is a binary Markov Random Fields model and encourages the
formation of clusters of like-valued binary variables.

The hyperparameters $\mathbf{a}$ control the sparsity of $\bgamma$. Since
we are focused on 2D and 3D lattices, which are regular graphs (i.e.,
each vertex has the same degree), we do not want to favor {a priori} the inclusion of any voxel. This is achieved by letting $\mathbf
{a} = a\mathbf{1}_p$, where $\mathbf{1}_p =(1,1,\dots,1)' \in\Re^p$. The
hyperparameters $\{b_{j_1,j_2}\}$ represent the prior belief on the
strength of coupling between the pairs of neighbors $(j_1,j_2)$, and
thus control the smoothness of $\bgamma$ over $\mathcal{E}$ given
$\mathbf{a}$, with larger $b_{j_1,j_2}$ leading to tighter coupling. When
$\bB=0$, the prior is the standard i.i.d. Bernoulli for each predictor
[\citet{GeoMcC93}]. Without specific prior information of
the strength of connection between each pair of neighbors, it is
natural to assume $b_{j_1,j_2}$'s to be a constant $b$. Then $(\mathbf{a},
\bB)$ reduce to two hyperparameters $(a,b)$, which can be either
pre-fixed or assumed to follow some hyperprior distributions.

The Ising prior smoothes the binary selection indicators, but not the
regression coefficients. In structured high-dimensional settings like
fMRI, neighboring covariates, often highly correlated, tend to have
similar effects on the outcome. Intuitively, a certain degree of
smoothing or grouping of the coefficients would improve the model
fitting, especially when the effects of individual predictors are very
weak. We achieve this by imposing a DP prior on $G$, $G \sim \operatorname{DP}(\alpha, G_0)$, with a precision parameter $\alpha$ and base measure $G_0$
[\citeauthor{Fer73} (\citeyear{Fer73,Fer74}); \citet{Ant74}].
Following the sticking-breaking (SB) presentation [\citet{Set94}],
$G$ can be written as a weighted sum of an infinite number of point
masses (atoms):
\begin{eqnarray}\label{SB}
G(\cdot) &=& \sum_{h=1}^{\infty}w_h\delta_{\theta_h}(\cdot), \qquad \theta_h \stackrel{\mathrm{i.i.d.}} \sim
G_0,
\nonumber\\[-8pt]\\[-8pt]\nonumber
w_h &=& w_h' \prod _{k<h} \bigl(1-w_k'\bigr),\qquad
w_h' \stackrel{\mathrm{i.i.d.}}\sim\operatorname{Beta}(1,\alpha),
\end{eqnarray}
where $\delta_\theta$ is a point mass at $\theta$. It is clear from
(\ref{SB}) that samples from a DP are discrete and the component
weights $w_h$ decrease exponentially in expectation. The spike-and-slab
prior (\ref{betaprior}) for each $\eta$ can then be written as a
mixture of an infinite number of point masses (at 0 and atoms randomly
drawn from the base measure $G_0$):
\begin{equation}
\label{SB-SSprior} \eta_j\mid (\gamma_j,\mathbf{w},\bolds{\theta})
\sim(1-\gamma_j)\delta_0+ \gamma_j \sum
_{h=1}^{\infty}w_h
\delta_{\theta_h}(\cdot),
\end{equation}
where $\bolds{\theta}=(\theta_1,\ldots,\theta_h,\ldots)$ and $\mathbf
{w}=(w_1,\ldots,w_h,\ldots)$. The clustering nature of the DP prior
can be immediately seen from (\ref{SB-SSprior}): it classifies the
voxels into one cluster of voxels that have no effect on response, and
several clusters of the remaining voxels, where the regression
coefficients within each cluster are shrunk to be identical. The number
of clusters increases automatically as the number of voxels under
consideration, $p$, increases. The precision parameter $\alpha$
governs the number of active components and is assumed to follow a
flexible hyper $\operatorname{Gamma}(1,1)$ prior. And we assume the base measure $G_0
=\normal(0,v^2)$ with hyperparameter $v$. In this article, clustering
{per se} is not the primary interest, rather clustering is a
means of grouping similar coefficients. There is a clear scientific
justification for grouping regression coefficients in this manner, as
each predictive brain region usually contains a number of voxels that
are of similar (and usually weak) effects on the outcome. Clustering
also introduces substantial improvement in posterior computation
because instead of sampling the coefficient for each voxel, one only
need to sample the common coefficient for each cluster.

Jointly, equations (\ref{isingprior}), (\ref{SB}) and (\ref
{SB-SSprior}) define the new \emph{Ising-DP spike-and-slab prior}.

\section{Selection of hyperparameters} \label{secparselection}
Selection of the hyperparameters $a,b$ in the Ising prior is crucial
for both inference and computational feasibility for high-dimensional
data. A challenging feature of the Ising prior in the ``large $p$''
paradigm is the phase transition behavior in a graph with dimension
higher than 1: certain combinations of the hyperparameters $a,b$ lead
to the selection of almost all variables and thus induce critical
slowdown of the MCMC for posterior computation. This issue cannot be
mitigated by simply replacing $a$ and $b$ by a hyperprior, because for
a regular graph with even modest degree (say, 3), the range of
hyperparameters that do not incur phase transition is narrow. If the
domain of the prior is not carefully chosen, it is very likely that
little weight is assigned to appropriate hyperparameters, leading to
poor posterior results, especially for data with low signal-to-noise
ratio (SNR), such as fMRI data. \citet{SmiFah07} suggested to
co-estimate the hyperparameters and the binary indicators in posterior
computation. Their method relies on specifying a uniform prior between
zero and a prespecified maximum for the smoothing parameter $b$.
However, if the maximum is specified outside the phase transition
bounds, the resulting MCMC will still suffer from the critical
slowdown. Therefore, finding these phase transition bounds is central
to correct specification of hyperparameters for the Ising prior.

Solely based on the prior distribution, \citeauthor{LiZha10} (\citeyear{LiZha10}), page 1205,
used mean field approximations to derive a ballpark estimate of the
phase transition boundary for the Ising prior defined on regular
graphs, and illustrated it on a hypertube with degree of 6. However,
because this approach does not take into account the data or any prior
knowledge of selection rate, it often results in a very wide range of
hyperparameters. The problem becomes even more pronounced when the
degree of the graph increases. Below we develop a new method to tighten
the bounds on $a$ and $b$ based on the posterior distribution.

The posterior conditional density of $\bgamma$ given the rest of
parameters is proportional to
\[
\mathcal{C}(\bgamma)=\exp \Biggl(\mathbf{a}'\bgamma+
\bgamma' \bB \bgamma-\sum_{i=1}^n
\bigl(Y_i-\bX_i(\bbeta\cdot\bgamma)\bigr)^2/2
\sigma ^2 \Biggr).
\]
In high-dimensional settings, usually it is reasonable to {a
priori} assume sparsity, that is, the proportion of true predictors
among the $p$ candidates, $\pi$, is much smaller than 1. Intuitively,
in order to have only a small proportion of predictors being selected,
the mode of $\mathcal{C}(\bgamma)$ should be larger than $\mathcal
{C}(\mathbf{0}_p)$ and attained at a $\hat{\bgamma}$ such that the
number of nonzero $\gamma$'s is around $\pi\cdot p$, beyond which
$\mathcal{C}(\bgamma)$ should decrease fast as the number of nonzero
$\gamma$'s increases. Below we form inequalities for $a$ and $b$ based
on this intuition.

When all the candidate voxels locate on a lattice, selected voxels give
rise to the largest number of neighboring pairs when they form a square
in two dimensions or a cube in three dimensions. Therefore, we use
squares (2D) or cubes (3D) to approximate the location of the selected
$\pi\cdot p$ voxels on the lattice. Let $V=[(\pi\cdot p)^{1/d}]$,
where $[c]$ denotes the largest integer no larger than $c$, and $d$ is
the dimension of the lattice, which equals either 2 or 3. For a square
containing $V^2$ voxels, there are $4V^2-6V+2$ neighboring pairs; for a
cubic containing $V^3$ voxels, there are $13V^3+28+66(V-2)+51(V-2)^2$
neighboring pairs (derivations are given in Appendix~\ref{appA}).

\subsection{Selection on two-dimensional lattice}
We first discuss the two-di\-men\-sional lattice. For $V^2$ selected voxels
on a square,
\[
\mathbf{a}'\bgamma+ \bgamma' \bB\bgamma=(a+8b)V^2-12bV+4b.
\]
To achieve sparsity, this value needs to decrease fast as $V$
increases, thus we must have $a+8b<0$. We also need the conditional
density of selecting $V^2$ voxels to be larger than the null model with
zero voxel, that is,
\begin{eqnarray}\label{ineq2}
&& -\sum_{i=1}^n(Y_i-
\bar{Y})^2/2\sigma^2
\nonumber\\[-8pt]\\[-8pt]\nonumber
&&\qquad \leq(a+8b)V^2-12 b V+4b- \sum_{i=1}^n\bigl(Y_i-
\bX_i(\bbeta\cdot\bgamma)\bigr)^2/2\sigma^2.
\end{eqnarray}
Since\vspace*{1pt} $\sum_{i=1}^n(Y_i-\bar{Y})^2$ is the total variation of the
observed $Y$, $\sum_{i=1}^n(Y_i-\bX_i(\bbeta\cdot\bgamma))^2$ is
the sum of squared errors, and $\bE\sum_{i=1}^n(Y_i-\bX_i(\bbeta
\cdot\bgamma))^2\approx n\sigma^2$, then $\sum_{i=1}^n(Y_i-\bar
{Y})^2/2\sigma^2-\sum_{i=1}^n(Y_i-\bX_i(\bbeta\cdot\bgamma
))^2/2\sigma^2\approx n\cdot\frac{R^2}{2(1-R^2)}$, where $R^2$ is
the determinant of coefficient in the linear regression of $Y$ versus
$\mathbf{X}$. Then inequality (\ref{ineq2}) is reduced to
\[
(a+8b)V^2-12 b V+4b>\frac{-n\cdot R^2}{2(1-R^2)}.
\]
We now propose two ways to determine $R^2$ to further tighten the
inequality. In the first method, we prespecify the $R^2$ value that we
expect to achieve. Then given $V$ from prior knowledge, obtain bounds
on the parameters $a$ and $b$. For example, if we want at least 50\% of
variation of $Y$ to be explained by the regression, and at most 5\% of
1000 voxels to be selected, we may let $R^2=50\%$ and $V=[\sqrt
{50}]=7$, then the inequality becomes $49(a+8b)-84b+4b>-n/2$, that is,
$312b+49a>-n/2$. Consequently, the range of $a$ and $b$ is determined
by two inequalities: $-8b>a>(-n/2-312b)/49$ and $b<n/160$. The second
method is to approximate $R^2$ by a lower bound obtained based on the
data: the maximum $R^2$ among all simple linear regressions of $Y$
versus each single predictor $X$. We believe such a lower bound is an
effective approximation for the problem under study for two reasons.
First, by using the DP prior, usually most of the selected voxels
should have identical $\beta$, effectively converting the multiple
regression to a simple linear regression. Second, for fMRI data,
spatially close voxels typically have very similar $X$ values, and thus
the $R^2$ value from regressing $Y$ on multiple spatially close
predictors is expected to be very similar to that from regressing $Y$
versus a single predictor.

\subsection{Selection on three-dimensional lattice}
Analogously, we can derive the range of $a$ and $b$ for a
three-dimensional lattice. For $V^3$ voxels forming a cubic and $V>1$,
\begin{eqnarray}\label{eq3d}
\mathbf{a}'\bgamma+ \bgamma' \bB\bgamma& = & (a+26b)
(V-2)^3+6(a+17b) (V-2)^2
\nonumber\\[-8pt]\\[-8pt]\nonumber
&& {}+12(a+11b) (V-2)+8a+56b.
\nonumber
\end{eqnarray}
In order to avoid all predictors being selected, we need $C(\bgamma
)<0$ to decrease fast as $V$ increases after certain threshold. For
simplicity, we only require $\mathcal{C}(\bgamma)$ to be negative for
the maximum possible $V$, that is, $V=[p^{1/3}]$. For example, in the
KLIFF data, $p$ is around 6600 in both ROIs, then $V=18$ and,
consequently, $a<-23b$.
In addition, in order to avoid the null model, that is, no voxel being
selected, we have
\begin{equation}
\label{Dim3InEq2} \mathbf{a}'\bgamma+ \bgamma' \bB\bgamma\geq
\frac{-n\cdot R^2}{2(1-R^2)}.
\end{equation}
Given the prespecified $R^2$ and $V$, we can obtain the range of $a$
and $b$ satisfying this inequality. Again taking the KLIFF data, for
example, $n=104$, we want at most $1\%$ voxels selected, and the
expected $R^2$ is $0.5$. Then $V=[66.7^{1/3}]=4$, plug this value and
$R^2=0.5$ into the inequality (\ref{Dim3InEq2}), and we have $a>-14.6b
-0.81$. Combining the previously obtained inequality $a<-23b$, it must
be the case that $-23b>-14.6b-0.81$, so that we have $b<0.1$.
Therefore, for the KLIFF data analysis, we will choose $a$ and $b$ such
that $b\leq0.1$ and $-23b>a>-14.6b-0.81$.

One potential problem of using (\ref{eq3d}) to evaluate $\mathbf
{a}'\bgamma+ \bgamma' \bB\bgamma$ in (\ref{Dim3InEq2}) is the
overestimation of the number of neighboring pairs of selected voxels,
especially when the selected $V$ is larger than 3, which can lead to a
very tight range of $b$ and $a$. We instead propose that as long as
there is one predictor whose posterior probability of being selected is
larger than that of not selected, the posterior simulation will not be
stuck at the null model. Therefore,\vspace*{-2pt} we can just let $a\geq\frac
{-n\cdot R^2}{2(1-R^2)}$, implying $b<\frac{n\cdot R^2}{2\cdot
23(1-R^2)}$ such that $-23b>\frac{-n\cdot R^2}{2(1-R^2)}$. For one of
the real data sets under study, the maximum $R^2$ across all simple
linear regressions is 0.10, then we have $-23b>a>-5.8$ and $b<0.25$.
Given the derived range of hyperparameters, and with the belief that
all the true predictors are tightly clustered together, we first choose
the largest possible $b$ to induce the most spatial clustering effect;
then given the value $b$, we choose the smallest $a$ within the phase
transition boundary to induce sparsity. Such a choice of $a$ also
brings computational advantage, because the computational cost of
obtaining the regression coefficients decreases with the number of
selected predictors in each MCMC iteration. Here, we choose $b=0.2$ and
$a=-4.5$ as the hyperparameters for the Ising prior.

\subsection{Remarks}
The above derivation suggests the following: first, the larger $R^2$
and the sample size $n$, and the smaller the degree of the underlying
graph (i.e., the average number of neighbors of each candidate
predictor), the wider the range of $b$; and second, the range of $a$
depends on both $b$ and the degree of the graph. Generally, for an
Ising model built on a regular graph, given $b$, a larger degree of the
graph leads to smaller $a$. These are consistent with a general
understanding of the effect of prior distributions in Bayesian
inference: when $R^2$ and $n$ are large, indicating a strong SNR and
abundant data information, choice of prior is less crucial. On the
other hand, if each predictor has many neighbors, then the positive
part $\bgamma^\prime\mathbf{B}\bgamma$ in the prior will give a
strong preference to models with many spatially close predictors.
Therefore, we need to use a smaller $b$ in order not to impose a strong
prior. This also explains, for fixed $b$, the larger the degree of the
graph, the smaller $a$ is required to induce a small prior odds of
selecting a large number of predictors.

The degrees of a 2D and 3D lattice are 8 and 26, respectively.
Consequently, the range of hyperparameters $a$ and $b$ that avoids
phase transition is much tighter in the latter than the former case.
Indeed, in the real application, when we assume the Ising prior on a 3D
lattice, the results are much more sensitive to the choice of $a$ and
$b$. In general, we find a larger degree of the underlying graph
corresponds to substantially more difficult hyperparameter selection
and inference, consistent with the observation made in \citet{LiZha10}. Also, it is crucial to examine $\bgamma^\prime\mathbf
{B}\bgamma$. Nevertheless, when choosing the underlying graph, the
concern of the degree of the graph should not outweigh the true
physical structure. For example, in fMRI data, we prefer an Ising prior
on a 3D lattice than on a 2D lattice, as the latter only accounts for
the structure in one slice and ignores the true 3D structure between voxels.

In Bayesian variable selection problems, choice of hyperparameters
affects not only posterior selection probabilities, but also
computational time, convergence rate and required iteration of MCMC
simulations. We found that if very few predictors are selected in each
iteration, the DP prior tends to shrink the $\bolds{\beta}$'s of all
predictors into one identical value, leading to very sticky MCMC, which
offsets the computational advantage per iteration offered by the
shrinkage effect of the DP prior. Therefore, besides avoiding the two
extreme ends of full selection and zero selection, the trade-off
between computation per iteration and convergence rate should be taken
into consideration when choosing the hyperparameters.

\section{Posterior computation} \label{secposterior}
We use a Gibbs sampler with data augmentation to carry out the
posterior inference of the proposed model: $\bgamma|-$, $\bbeta|-$, $\sigma
|-$, where ``$-$'' denotes all the rest of the parameters. Below we
describe the outline of the Gibbs sampler but relegate the
computational details to Appendix~\ref{appB}.

The procedure to update the variance $\sigma$, and the indicators
$\bgamma$, which we update one at a time in a random order in each
sweep, is standard. To draw posterior samples of $\bbeta$, we use an
approximate blocked Gibbs sampler based on the truncated stick-breaking
process [\citet{IshZar00}; \citet{IshJam01}].
First choose a conservative upper bound, $H < \infty$ on the number of
mixture components potentially occupied by $\beta_j$'s in the sample.
Then introduce latent class indicators for each predictor, $Z_j  (\in
\{1, \ldots, H\})$ with a multinomial distribution, $Z_j \sim\mathrm{MN}(\bw)$ where $\bw=\{w_1, \ldots, w_H\}$. This associates each
predictor in the current iteration with a cluster $h$ in the DP. In the
Gibbs sampler, we first augment the cluster membership $Z_j$ and then
sample $\beta_j$ conditional on $Z_j$.

The main computational gain, especially when $p$ is large, is due to
the clustering nature of DP: because all the predictors in one cluster
share the same coefficient, we only need to update one $\beta$ for
each cluster within each iteration. It is easy to show the
computational order of the posterior computation of one MCMC iteration
under the DP prior for $\beta$ is $O(n\times p\times p_{\mathrm{sel}})$, where
$p_{\mathrm{sel}}$ is the number of selected predictors (model size) in that
iteration. For comparison, we present the corresponding computational
order under the standard spike-and-slab prior with Gaussian prior for
$\beta$, for which there are two general schemes for posterior
computation: (i) sample all parameters, $\beta$, $\sigma$ and $\gamma
$; and (ii) integrate out $\beta$ and $\sigma$ under the conjugate
setup and only sample $\gamma$. In both schemes, the main
computational burden is due to the inversion of the covariance matrix,
which, even using fast low-rank update algorithms, is of the order
$O(n\times p^2)$ and $O(n\times p\times p_{\mathrm{sel}}^2)$, respectively. When
$p$ is very large as in this application, the computational order of
the first scheme is prohibitive, and this is the reason that the vast
majority of the SSVS literature in high-dimensional settings adopts the
second scheme, which, however, does not provide posterior samples of
the coefficients $\beta$ or the variance $\sigma$. Moreover, because
of the squared term of $p_{\mathrm{sel}}$, even when the average model size is
modest (e.g., between 50--100), the second scheme can still incur
overwhelming computational cost. In contrast, as shown in the details
of the Gibbs sampler in Appendices~\ref{appA} and \ref{appB}, the DP prior does not require
matrix inversion, yet still provides posterior samples of $\beta$'s
with much lower computational cost.

\section{Simulations} \label{secsimulation}
\subsection{Simulation design}
We conduct simulations to examine the performance of the Ising-DP prior
and compare with several alternative methods. We simulate data of
$n=104$ subjects (the number of subjects in the real application), each
having $p=1000$ candidate predictors overlaying a $10\times10\times
10$ 3D grid. Each predictor $j$ ($1\leq j\leq1000$) is spatially\vspace*{1pt}
indexed by $\mathbf{d}_j=(d_j^1,d^2_j,d^3_j)$ for $1 \leq
d_j^1,d_j^2,d_j^3\leq10$. To mimic the real data, we let predictors be
strongly correlated, and the design matrices of the $i$th subject
$\mathbf{X}_i=(X_{i1},\ldots, X_{ip})$ in all the following simulations
follow a multivariate normal $\mathrm{MVN}_p(\bolds{\mu},\Sigma)$, where
$\bolds{\mu}=(\mu_1,\ldots,\mu_p)\stackrel{\mathrm{i.i.d.}}{\sim}\operatorname{Unif}(3,6)$ and $\Sigma_{j_1j_2}=0.8^{\llvert  \mathbf{d}_{j_1}-\mathbf
{d}_{j_2}\rrvert  }$, where $\llvert  \mathbf{d}_{j_1}-\mathbf{d}_{j_2}\rrvert  =\sum_{i=1}^3
\llvert  d_{j_1}^i-d^i_{j_2}\rrvert  $. We consider the following four simulation scenarios.
\begin{longlist}[\textit{Scenario} 1:]
\item[\textit{Scenario} 1:] \textit{One cluster of true predictors}, \textit{with identical
$\beta$}'\textit{s}. There is a cluster of $5\times5\times5$ (125) true
predictors ($\gamma_j=1$) with spatial indices $4 \leq
d_j^1,d_j^2,d_j^3\leq8$ located in the center of the 3D cube. The
coefficients $\beta$ of the true predictors are set to 0.6. The
response is generated from $Y_i=\sum_j X_{i,j}\beta_j\gamma
_j+\varepsilon_i$ with $\varepsilon_i \sim\normal(0,200^2)$ for
$i=1,\ldots,n$, creating a data set with a low SNR 5\%---defined as $\bV
(\mathbf{X}\bolds{\beta})/\bV(\bolds{\varepsilon})$. The following
scenarios also all have such a low SNR, which is the norm in real fMRI data.

\item[\textit{Scenario}  2:]  \textit{One  cluster  of  true  predictors},
\textit{with  varying  but strongly  correlated}  $\beta$'\textit{s}. We let the coefficients of the true
predictors, locating on the same grid as those in scenario~1, vary and
follow $\mathrm{MVN}_p(0.6\times\one_p,\Omega)$, where $\Omega
_{j_1j_2}=0.1\times0.95^{\llvert  \mathbf{d}_{j_1}-\mathbf{d}_{j_2}\rrvert  }$. Therefore,
both the observed values and the underlying coefficients of neighboring
predictors are strongly correlated.

\item[\textit{Scenario} 3:] \textit{Two clusters of true predictors}, \textit{with identical}
$\beta$'\textit{s within each cluster}. A more challenging scenario is when
there are multiple spatially separated clusters of true predictors.
Specifically, we let the true predictors form two clusters: one
overlays the grid of $3 \leq d_j^1\leq4,3\leq d_j^2\leq4,3\leq
d_j^3\leq4$, and another overlays the grid of $6 \leq d_j^1\leq
9,6\leq d_j^2\leq9,6\leq d_j^3\leq9$. We set the coefficients $\beta
$ of the predictors in the two clusters to 0.4 and 1, respectively.

\item[\textit{Scenario} 4:] \textit{Two clusters of true predictors, with varying} $\beta
$'\textit{s within each cluster}. The true predictors locate on the same grid
as those in scenario~3, and one cluster of $\bolds{\beta}$ were
generated from $\mathrm{MVN}_p(0.4\times\one_p,\Omega_1)$ with
$\Omega_{1,j_1j_2}=0.1\times0.95^{\llvert  \mathbf{d}_{j_1}-\mathbf{d}_{j_2}\rrvert  }$,
and those in the second cluster are from $\mathrm{MVN}_p(1\times\one
_p,\Omega_2)$ with $\Omega_{2,j_1j_2}=0.1\times0.95^{\llvert  \mathbf
{d}_{j_1}-\mathbf{d}_{j_2}\rrvert  }$. Variable selection under two-cluster
scenarios is challenging: the strong correlation between the predictors
outside and inside the clusters renders differentiating nonsignificant
predictors, especially those located between the two clusters, from the
true ones difficult.

For each of the simulated data set, we fit the regression model (\ref
{reg}) with four different priors: (i) i.i.d. Bernoulli prior for
$\gamma_j$, with a Gaussian prior for the $\beta_j$'s (this is the
standard spike-and-slab prior), referred to as the \emph{i.i.d.-Gaussian}
prior; (ii) Ising prior for $\gamma_j$, with a Gaussian prior for the
$\beta_j$'s, referred to as the \emph{Ising-Gaussian} prior; (iii)
i.i.d. Bernoulli prior for $\gamma_j$, with a DP prior for $\beta
_j$'s, referred to as the \emph{i.i.d.-DP} prior; (iv) the \emph
{Ising-DP} prior. The hyperparameters $(a,b)$ for the Ising priors are
chosen by the proposed approach in Section~\ref{secparselection},
with $a=-5$ and $b=0.25$. For the DP priors, we set $H=20, \alpha=1$
and $v=10$ such that $G_0$ is very flat in a wide domain. For each
simulated data, we run 10 parallel Gibbs samplers with random start in
$\bgamma$, each having 20,000 iterations with the first 10,000 ones as
burn-in. Posterior computation with the i.i.d.-Gaussian and Ising-Gaussian
priors are carried out using the software by \citet{LiZha10}. The
main summary statistic, the posterior inclusion probability, is deemed
convergent upon inspecting the Gelman--Rubin statistic [\citet{GelRub92}]. In all of our experiments, the 10 simulations lead to highly
similar posterior summary statistics.
\end{longlist}

\subsection{Simulation results}
We calculate the posterior inclusion probabilities $\Pr(\gamma
_j=1\mid \bY)$ as the posterior summary statistics, obtained by dividing
the number of iterations where $\gamma_j=1$ over the total number of
iterations excluding the burn-in period. To summarize these marginal
probabilities, we compute the ROC curve as follows: only those
covariates $j$ with $\Pr(\gamma_j=1\mid \bY)$ greater than a threshold
are deemed positives, and those below the threshold are deemed
negatives; the ROC curve reflects the pair of true positive rate and
false positive rate achieved by varying the calling threshold. The
bigger area under the ROC curve (maximum 1), the better the
discriminating power of the model.

\begin{figure}
\begin{tabular}{@{}cc@{}}
\multicolumn{2}{@{}c@{}}{\footnotesize{One cluster}}\\[3pt]

\includegraphics{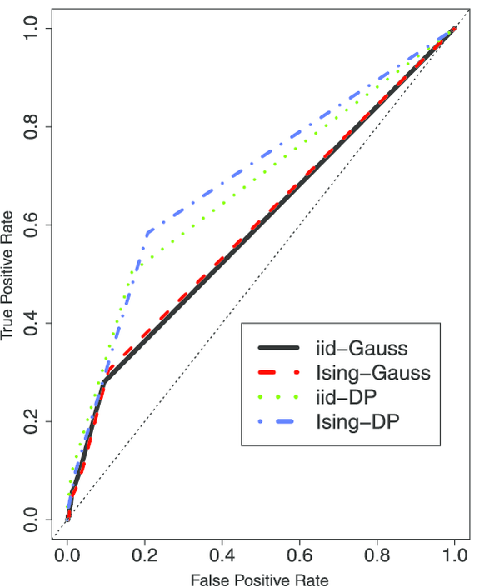}
 & \includegraphics{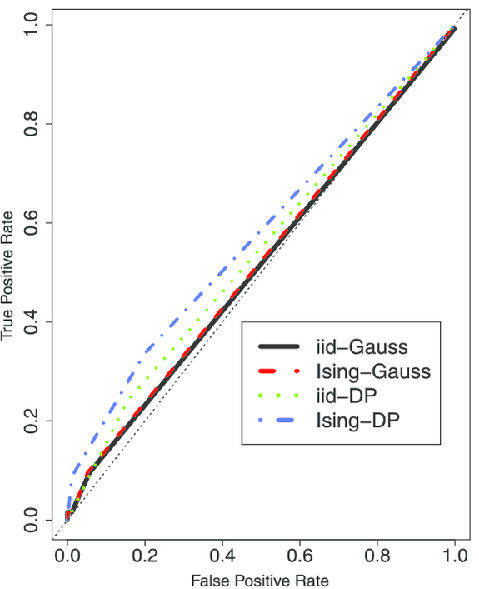}\\
\footnotesize{(a) Identical $\beta$ of true predictors} & \footnotesize{(b) Varying $\beta$ of true predictors}\\[6pt]
\multicolumn{2}{@{}c@{}}{\footnotesize{Two clusters}}\\[3pt]

\includegraphics{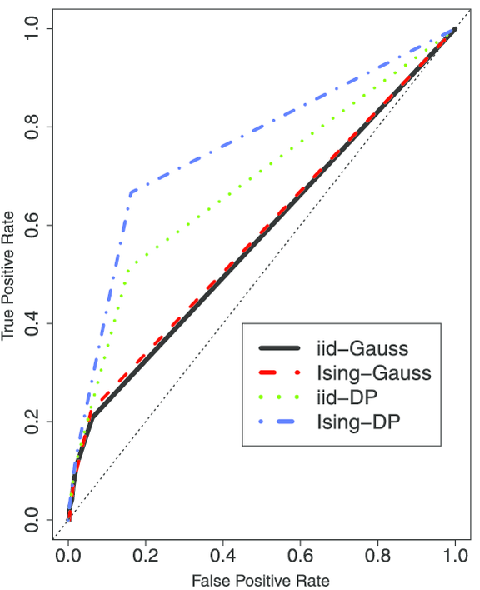}
 & \includegraphics{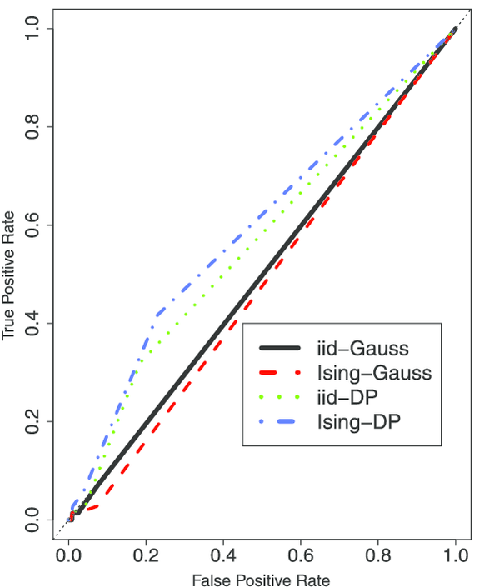}\\
\footnotesize{(c) Identical $\beta$ of true predictors within a cluster} & \footnotesize{(d) Varying $\beta$ of true predictors}
\end{tabular}
\caption{ROC curves based on the posterior selection probability $\Pr
(\gamma_j=1\mid \bY)$ obtained from i.i.d.-Gaussian, Ising-Gaussian, i.i.d.-DP
and Ising-DP prior, respectively, under four simulation scenarios.}\label{figROC}
\end{figure}

The ROC curves resulting from the simulations under scenarios 1--2 (one
cluster) and 3--4 (two clusters) are presented in the top and bottom
panel of Figure~\ref{figROC}, respectively. We also calculated the
root mean squared error (RMSE) per variable, $(\sum_j(\hat{\beta
_j}-\beta_j)^2/p)^{1/2}$, of each prior, summarized in Table~\ref{tabrmse}.
In all four simulations, the Ising-DP prior resulted in the best ROC,
closely followed by the \mbox{i.i.d.-DP} prior, beating both the i.i.d.-Gaussian and
the Ising-Gaussian priors. This pattern is consistent with the RMSEs.
Overall, the ROC curves suggest relatively low discriminating power in
these simulations, even for the best-performing Ising-DP prior. This is
not surprising because variable selection under all four scenarios is
very challenging due to the low SNR, strong correlation between
variables and the small-$n$ large-$p$ nature. Indeed, our experience
based on more simulations suggests that as the SNR and/or the sample
size decreases, performance of all the priors drops, but the Ising-DP
prior is the least affected, demonstrating the benefit of introducing
additional shrinkage to the coefficients when the signal is weak. In
summary, it is evident from these simulations that the Ising-DP prior
outperforms the existing alternatives in data with characteristics
similar to those of the fMRI data under study.

%

It is worth noting that in these simulations the DP component appears
to impose a stronger clustering effect on performance than the Ising
component. One reason is that, as shown in Section~\ref{secparselection}, when the degree of the graph is large as in the 3D
fMRI analysis, the hyperparameter $b$ in the Ising prior used to
control the clustering effect has to be set small to avoid phase
transition, which consequently limits its clustering effect.
Nevertheless, the simulation results suggest that incorporating the
spatial information into Bayesian variable selection via the Ising
prior still leads to improved selection accuracy than otherwise.

%
\begin{table}[t]
\tabcolsep=0pt
\caption{Root mean squared error (RMSE) per variable, $(\sum_i(\hat
{\beta_i}-\beta_i)^2/p)^{1/2}$, by different priors} \label{tabrmse}
\begin{tabular*}{\tablewidth}{@{\extracolsep{\fill}}@{}lcccc@{}}
\hline
\textbf{Scenario} & \textbf{I.i.d.-Gaussian} & \textbf{Ising-Gaussian} & \textbf{I.i.d.-DP} & \textbf{Ising-DP}\\
\hline
1. One-cluster identical $\beta$ &0.623 &0.599 &0.190 &0.190\\
2. One-cluster varying $\beta$ &0.284 &0.283 &0.181 &0.179 \\
3. Two-cluster identical $\beta$ &0.311 &0.315 &0.256 &0.250\\
4. Two-cluster varying $\beta$ &0.368 &0.251 &0.235 &0.233\\
\hline
\end{tabular*}
\end{table}

\section{Application to the KLIFF study} \label{secapplication}
\subsection{The data}
We now provide more information on the design of the KLIFF study and
the preprocessing procedure. For each of the 104 pairs of participants
in a close relationship (referred to as partners hereafter), one of
them was randomly selected to be threatened by electric shocks while
their brain activities were measured by fMRI in three separate
sessions: in one session he/she is holding hands with his/her partner;
in the second session, he/she is holding hands with a stranger; in the
third session, he/she is alone, holding hands with nobody at all. The
three hand-holding conditions mimic three types of social interactions.
Each of the three sessions, randomized within each pair of partners,
contains 24 trials in random order, half of which are threat cues (a
red ``X'' on a black background) indicating a 20\% likelihood of
receiving an electric shock to the ankle, and the other half are safety
cues (a blue ``O'' against a black background) indicating no chance of
shock. A 3D fMRI scan of the subject's brain was acquired for every 2
seconds in the experiment lasting for 400~s. Overall, fMRI data
collected from the KLIFF experiment consist of 104 subjects in 3
sessions at 200 time points for over 100,000 spatially distributed
voxels. At the end of each session, the subjects facing the threat were
asked to score their arousal and valence feelings experienced during
the experiment. Both the arousal and valence measurements range from 1
to 9, encoding feelings from calming/soothing to alert/agitated, and
feelings from highly negative/miserable to highly positive/pleased,
respectively.

Preprocessing of the fMRI data was carried out via FMRIB's Software
Library (FSL) software [Version 5.98; \citet{Smietal04}].
Registration of the images in FLIRT [\citet{Jenetal02}] was based
on Montreal Neurological Institute (MNI) space. More details of
preprocessing can be found in \citet{Zhaetal13}. ROIs were determined
structurally using the Harvard subcortical brain atlas, and were chosen
for their likely involvement in affective processing based on previous
studies [\citet{MarBecCoa13}]. In particular, our analysis focuses on
two emotion related regions: dorsal anterior cingulate cortex (dACC)
and insula, which were commonly implicated in negative affect and
threat responding, and whose numbers of voxels are similar, 6666 and
6591, respectively. To obtain the predictors, we conducted massive
univariate analysis using the GLM to get scalar summaries of the fMRI
time series. Specifically, for every voxel in each ROI, we \textit{used} the semi-parametric GLM approach in \citet{Zhaetal13} to
estimate the hemodynamic response functions (HRF) corresponding to the
threat and safety cues (stimuli), and extracted the height of the HRF
estimates, interpreted as the magnitude of brain response to the
stimuli of that voxel. We then computed the difference between the
estimated magnitudes under the threat cue and the safety cue (baseline)
for each voxel as the predictors. In total, for each ROI, we obtained
six sets of regression data: two different response variables---valence and arousal scores of the subjects, under each of the three
hand-holding conditions, and associated magnitude estimates of each
voxel in the ROI collected in the same session as the predictors.

\subsection{Results}
We applied the proposed Bayesian model to the 12 sets of data (6 for
each ROI) using the Ising-DP prior on a 3D lattice with hyperparameter
$a=-4.5$ and $b=0.2$ obtained from the method in Section~\ref{secparselection}. For comparison, we also fit the model with the
i.i.d.-Gaussian and the Ising-Gaussian priors. For each regression, 25,000
iterations of MCMC were performed with the first 5000 discarded as
burn-in. Convergence of the marginal inclusion probabilities is deemed
via the Gelman--Rubin statistics.

Though the number of selected predictors is larger than the sample size
in each MCMC iteration, the clustering effect of the DP prior leads to
a small number of different $\beta$ values (less than 10) in most
iterations. Among the 12 sets of regressions, we focused on those with
(i) reasonably high R-squared values and (ii)~top 10\% selected voxels
having a high proportion of nonzero coefficients with the same sign.
The R-squared value for each iteration $t$ is given by
\[
R^2_t =1-\Var(\bY-\bX \bolds{\gamma}_t\cdot
\bolds{\beta}_t)/\Var (\bY),
\]
where $\bolds{\gamma}_t$ and $\bolds{\beta}_t$ are the posterior draws of
$\bolds{\gamma}$ and $\bolds{\beta}$, respectively, at the $t$th
iteration. The first criterion requires that a significant proportion
of variation of subjects' emotion measurements can be explained by
their brain response magnitudes, and the second requires that the
majority of the top selected predictors have similar and significant
effects on the response, matching the substantive knowledge from the
existing psychology literature. We found three sets of regressions fit
these two criteria: the regression with the arousal measurement under
alone condition as the response in dACC and insula, respectively, and
the regression with the valence measurement under
hand-holding-with-partner condition as the response in insula.

%
\begin{figure}[t]
\begin{tabular}{@{}ccc@{}}

\includegraphics{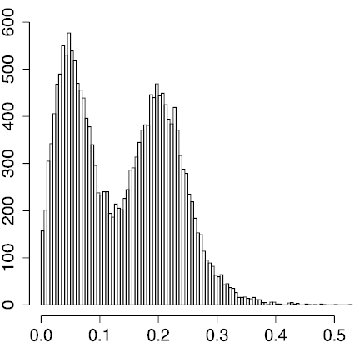}
 & \includegraphics{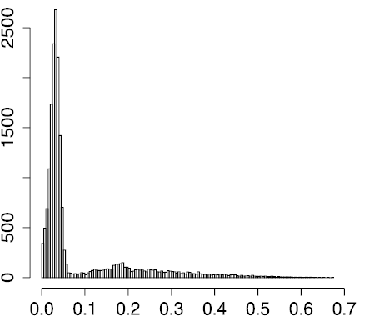} & \includegraphics{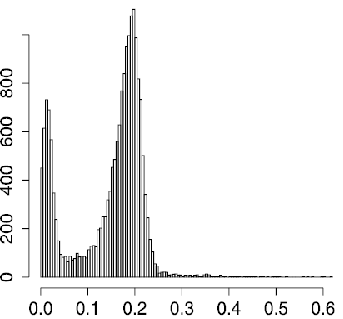}\\
\footnotesize{(a) dACC alone arousal} & \footnotesize{(b) Insula alone arousal} & \footnotesize{(c) Insula partner valence}
\end{tabular}
\caption{R-squared values of the regressions.}\label{figRsquare}
\end{figure}

%
\begin{figure}[b]
\begin{tabular}{@{}ccc@{}}

\includegraphics{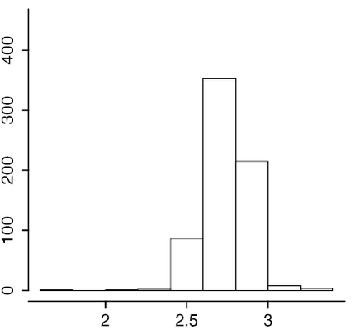}
 & \includegraphics{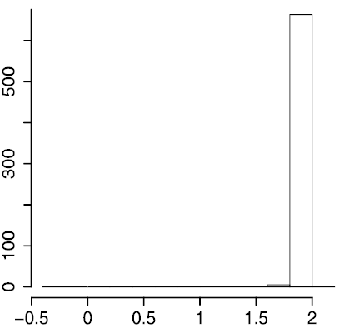} & \includegraphics{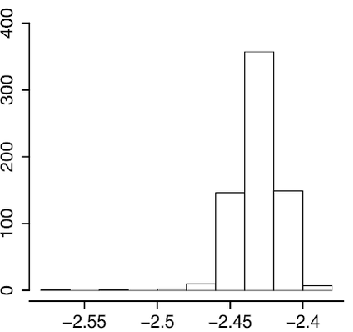}\\
\footnotesize{(a) 10\% percentile} & \footnotesize{(b) 10\% percentile} & \footnotesize{(c) 90\% percentile}\\[-2pt]
\footnotesize{dACC arousal} & \footnotesize{insula arousal} & \footnotesize{insula valence}
\end{tabular}
\caption{Histograms of 10\% or 90\% percentile of the coefficients (in
scale $10^{-4}$) of the top 10\% selected voxels in dACC and insula
when regressing subjects' arousal (the first two figures) or valence
(the third figure) scores versus the magnitude of brain response to
threat under the alone or hand-holding-with-partner condition.}
\label{figHistBeta}
\end{figure}

Histograms of the R-squared values and the coefficients of the top 10\%
selected voxels in these three regressions are displayed in Figures~\ref{figRsquare} and \ref{figHistBeta}, respectively. We can see
that in the regression with arousal under the alone condition as the
response in dACC, the R-squared value is larger than 20\% in more than
20\% of the MCMC draws [Figure~\ref{figRsquare}(a)], and almost all
($>$99.5\%) of the top 10\% selected voxels' coefficients are positive
in more than 90\% of the posterior draws [Figure~\ref{figHistBeta}(a)]. The same regression in insula led to similar
results [R-squared in Figure~\ref{figRsquare}(b) and coefficients in
Figure~\ref{figHistBeta}(b)]. The significant positive association
between the arousal measurement and brain response magnitudes under the
alone condition is consistent with related findings in the literature.
First, in a previous study of the KLIFF data [\citet{Zhaetal13}], we
found that the brain response to threat stimulus is most active when
subjects are alone. This phenomenon can be explained through the social
baseline theory [\citet{BecCoa11}; \citet{CoaBecAll13}; \citet{CoaMar14}], which suggests that the human brain assumes proximity
to other human beings, and perceives the environment as less
threatening during the presence of other people in a close
relationship, and thus serving as a default, or baseline, strategy of
emotion regulation. This reduces the need to rely on effortful
self-regulation in response to threat. On the other hand, when the
subjects are alone without any social support, their brains have to use
their own energy for emotion regulation, and, consequently, their
emotional response is strong, and its association with subjects'
emotion measurements is easier to detect in the two emotion-related
ROIs. Second, the positive association between brain response and
excitement level corresponds with literature showing a role for dACC
and insula in both cognitively- and physically-induced arousal
[\citet{Crietal00}; \citet{Lewetal07}]. Since the use of
electric shock as a threat stimulus causes physical pain and induce
subjects' internal awareness of upcoming pain during anticipation of a
shock particularly, it is natural that the more active emotion-related
ROIs process the stimulus, the more intense and agitated feeling the
subjects experience.

%
\begin{figure}

\includegraphics{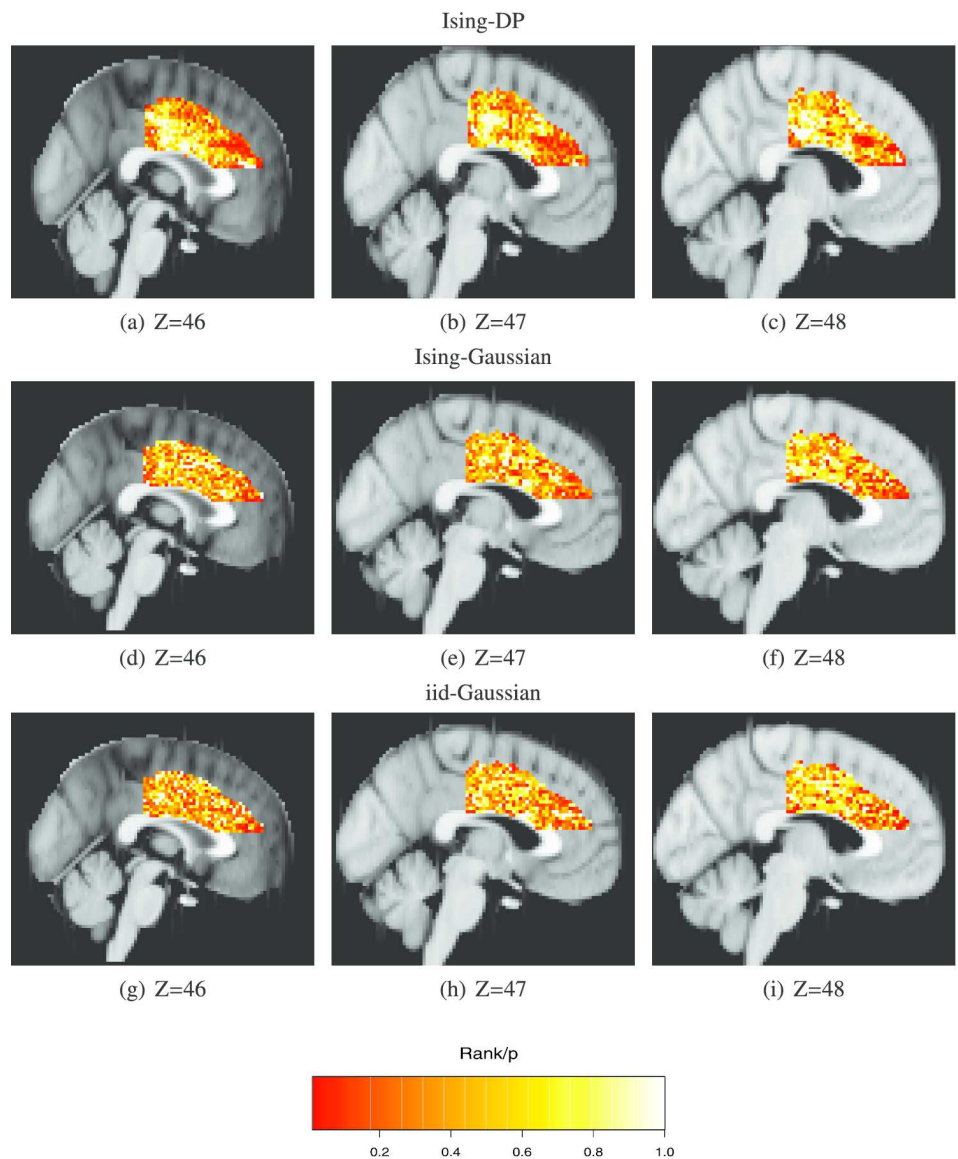}

\caption{Heatmaps of voxels according to the ranks of their posterior
inclusion probabilities obtained from Ising-DP, Ising-Gaussian and
i.i.d.-Gaussian priors, respectively, in the Bayesian regression of
subjects' arousal scores versus the magnitude of brain response to
threat of voxels in dACC and insula when subjects are alone.}\label{figAloneArousaldACC}
\end{figure}

%
\begin{figure}

\includegraphics{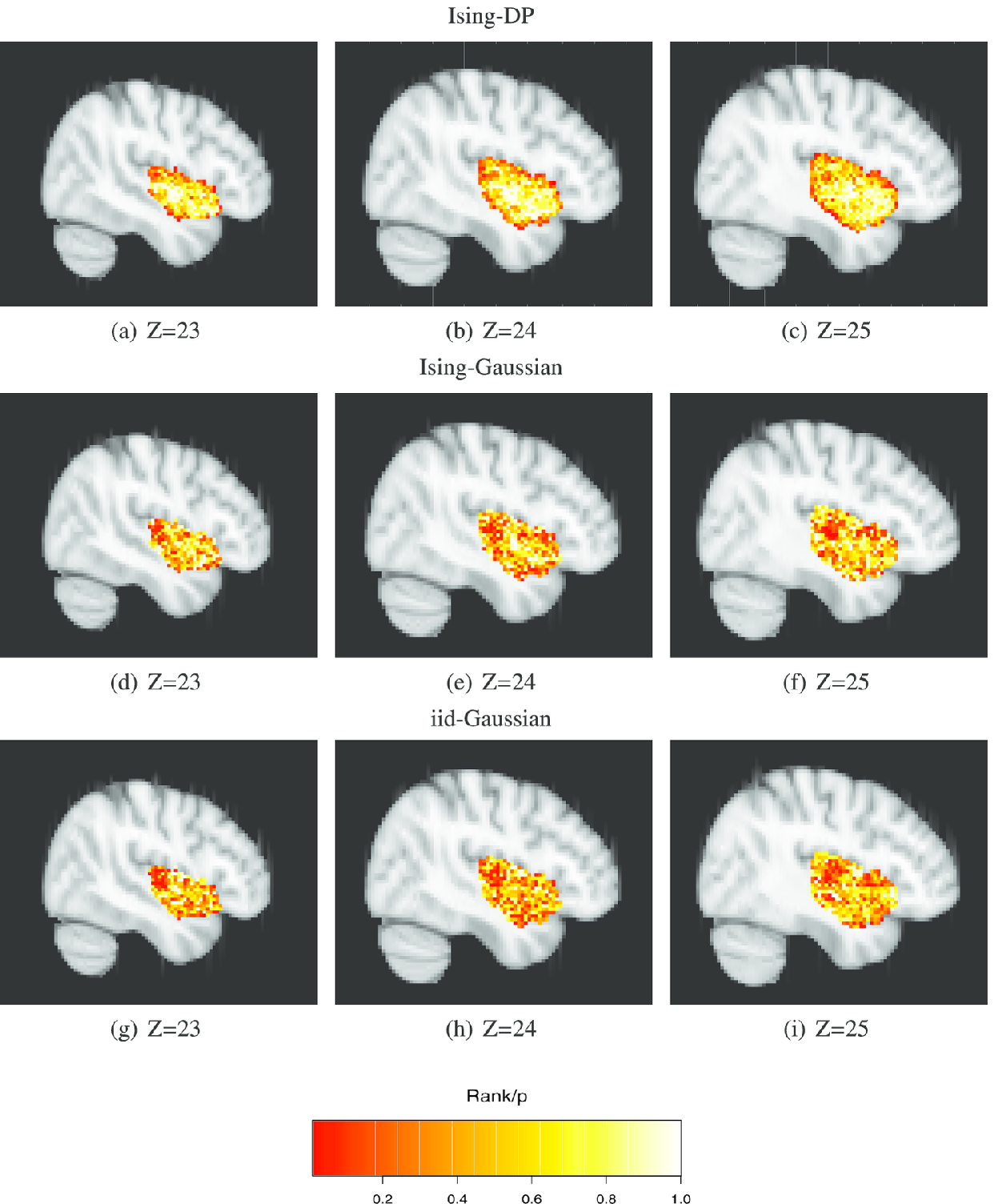}

\caption{Heatmaps of voxels according to the ranks of their posterior
inclusion probabilities obtained from Ising-DP, Ising-Gaussian and
i.i.d.-Gaussian priors, respectively, in the Bayesian regression of
subjects' arousal scores versus the magnitude of brain response to
threat of voxels in insula when subjects are alone.}\vspace*{-4pt}
\label{figAloneArousalInsula}
\end{figure}

%
\begin{figure}

\includegraphics{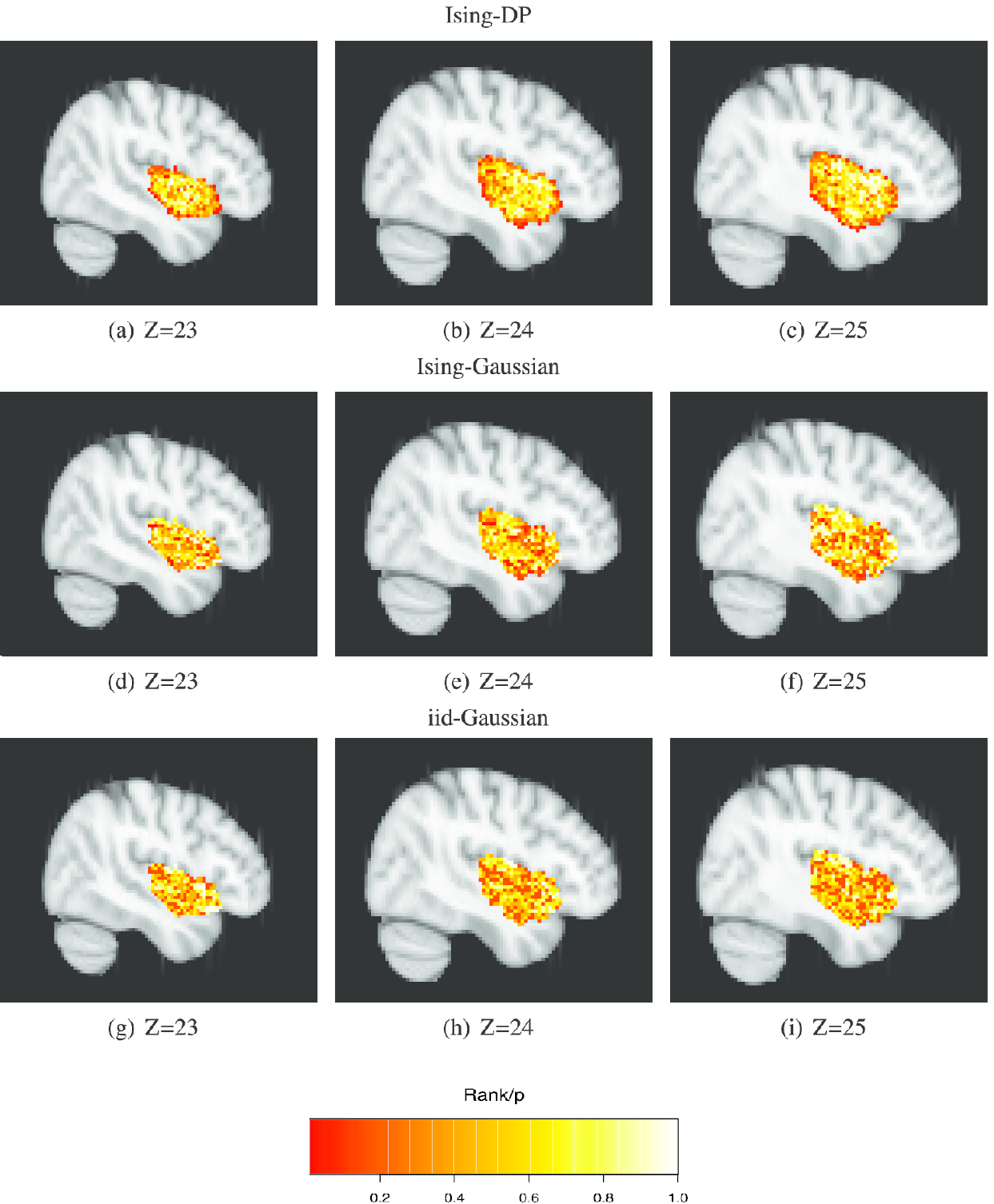}

\caption{Heatmaps of voxels according to the ranks of their posterior
inclusion probabilities obtained from Ising-DP, Ising-Gaussian and
i.i.d.-Gaussian priors, respectively, in the Bayesian regression of
subjects' valence scores versus the magnitude of brain response to
threat of voxels in insula when subjects are hand holding with their
partners.}\vspace*{-4pt}\label{figPartnerValenceInsula}
\end{figure}

We also found significant association between valence and brain
response magnitude in insula under hand-holding-with-partner condition
[R-squared values shown in Figure~\ref{figRsquare}(c) and
coefficients shown in Figure~\ref{figHistBeta}(c)]. The negative
association has two possible explanations. First, the threat stimulus
induces subjects' negative feelings, and the valence and arousal
measures are negatively correlated, therefore, the more active the
brain responds to the stimulus, the less pleased the subjects'
feelings. Second, according to the social baseline theory, humans feel
less threatened under the hand-holding-with-partner condition. Thus,
subjects' emotion variation is more likely to occur in the valence
dimension. We indeed found that the variance of subjects' valence is
larger than that of arousal. Moreover, insula is thought to mediate the
awareness of internal bodily and emotional states [\citet{Cra09}] and is
related to pain anticipation and intensity [\citet{WiePloTra08}].
Results of the regression under the hand-holding-with-stranger
condition are not as stable as the other two regressions, possibly due
to the individual differences in cognitive and affective perception of
strangers.

In all three regressions, the largest posterior selection probabilities
of voxels are around 0.1, and the majority of the probabilities are
below 0.05. This is as expected given the very low SNR common in fMRI
data. In these situations, arguably, the ranks rather than absolute
value of the probabilities are more informative about the selection
results. Figures~\ref{figAloneArousaldACC}, \ref
{figAloneArousalInsula} and \ref{figPartnerValenceInsula} show the
heatmaps of the posterior selection probabilities of the voxels in
three slices based on their rank under the Ising-DP (top panel) in
these regressions, respectively, in comparison to the corresponding
heatmaps under the i.i.d.-Gaussian (middle panel) and the Ising-Gaussian
prior (bottom panel). The color scale is arbitrary, with dark red
representing the selection probability in the lowest rank and light
yellow representing the highest rank. The most striking pattern from
these graphs is that the areas with the highest selection probabilities
identified by the Ising-DP prior were smoothly located across the ROIs,
matching the scientific understanding of human brain functions, in
contrast to those by the i.i.d.-Gaussian or the Ising-Gaussian prior,
which are very diffused and scattered across the entire region.

\begin{figure}
\begin{tabular}{@{}c@{\hspace*{3pt}}c@{\hspace*{3pt}}c@{}}

\includegraphics{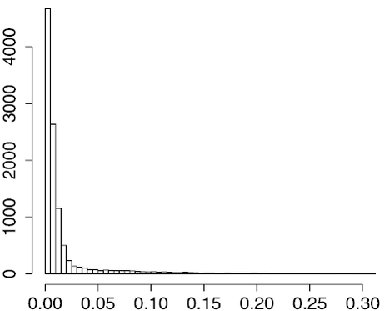}
 & \includegraphics{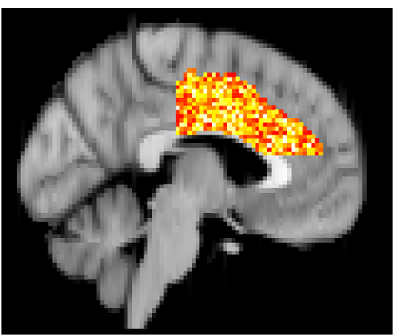} & \includegraphics{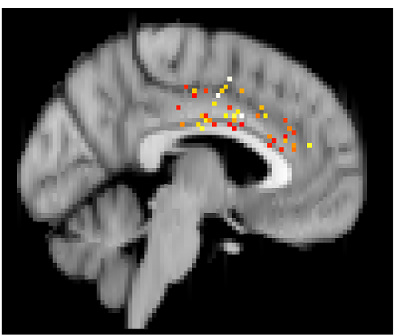}\\
\footnotesize{(a) Histogram of R-squared} & \footnotesize{(b) Heatmap of selection} & \footnotesize{(c) Voxels with highest top 10\%}\\[-2pt]
& \footnotesize{probabilities of voxels} & \footnotesize{selection probabilities}
\end{tabular}
\caption{Regression of simulated response versus brain activity
measurements in dACC under alone condition.}
\label{figRsquareNull}\label{figSimuYRsquareddACCAloneArousal}
\end{figure}

Since the underlying truth is unknown, we use a simulation-based
procedure to obtain the sampling distribution of the R-squared values
of a null model. Specifically, we simulated, independently of the
covariates, a normally distributed response variable with similar
variance and range as the observed emotion measurements, and applied
the Bayesian model to regress the simulated outcome on the observed
covariates in dACC under the alone condition. The histogram of positive
R-squared values in the posterior draws of this null model, shown in
Figure~\ref{figSimuYRsquareddACCAloneArousal}(a), centers around
zero, and is distinct from the histograms from the aforementioned three
regressions, each of which has a much higher proportion of large
R-squared values. In contrast, the histogram of the null model is very
similar to those from the remaining nine regressions. As such, we deem
there is no statistically significant association between the
covariates and the responses in these nine regressions.

%

\section{Discussion} \label{secconclusion}
Motivated by the KLIFF hand-holding experiment, in this article we
propose a joint Ising-DP prior within the Bayesian SSVS framework to
achieve selection and grouping of spatially correlated variables in
high-dimensional SI regression models. We developed an analytic
approach for deriving the bounds of the hyperparameters to avoid phase
transition, a main challenge in methods involving the Ising prior.
Though the bounds provided by our method are tighter than the previous
mean field bounds, they are still only ballpark estimates and may be
wide in graphs with high degrees. A focus of our future research is
therefore to improve the method of hyperparameter selection for a more
complex graphical structure.

A major challenge to MCMC-based Bayesian methods in high-dimensional
settings is computation. Though the DP prior in our model partially
reduces the computational load by clustering the coefficients,
computational scalability remains a challenge given the large $p$.
Indeed, currently we are not able to perform a whole brain analysis
with $p\approx{}$100,000. Moreover, the mixing rate of the MCMC of the
standard strategy in SSVS of updating one variable at a time may be
slow, especially when the DP prior is involved. An attractive direction
is to design a block update Gibbs sampling scheme that updates multiple
variables at a time, and to parallelize the computation within a block
using graphics processing unit (GPU)-based programming [\citet{Sucetal10}; \citet{Geetal14}]. The procedure can be further speeded up by
carefully selecting the block so that it matches the underlying block structure.

The Ising prior is a special case of Markov random fields. \citet{KalSamFah14}
proposed latent GMRFs via a probit model.
The probit-GMRF prior simplifies the calculation of the hyperparameters
and does not suffer from the phase transition behavior. However, the
main computational hurdle of inversion of a matrix of the size of
selected variables remains. Nevertheless, it is possible to combine the
DP prior with the probit-GMRF prior to reduce the computation.

Extension to binary and categorical responses is, in principle,
straightforward using generalized linear models. Computation is an
increased focus, as closed-form posterior conditional distributions are
no longer available. The same problem applies with censored survival
models. Laplace approximations [\citet{Raf96}] are useful, but they
usually require gradient methods for iterative computation of posterior
modes for each sweep of covariates. A possible improvement can be
obtained by exploiting the majorization--minimization/maximization (MM)
algorithm [\citet{Lan13}]---a generalized version of the EM
algorithm---for within-model mode computations.

The proposed Ising-DP prior inherently assumes sparsity, that is, only
a small portion of the voxels in the ROIs are associated with the
individual scalar outcome. This is achieved via a point mass
(spike-and-slab) prior for the regression coefficients, resulting in a
``hard-thresholding'' of the $\beta$'s. However, in our real
application, posterior probabilities of inclusion of nearly all voxels
are relatively small, which suggests that an alternative
``soft-thresholding'' without sparsity---achieved by (spatial adaption
of) LASSO-type priors [\citet{ParCas08}]---may be desirable and
a worthwhile direction for future investigation.

Though we have focused on fMRI, the proposed model is applicable to
other imaging modalities where detailed spatial information between
covariates is available, such as DTI or MRI.

Matlab code that implements the method is available at
\surl{http://faculty.\\virginia.edu/tingtingzhang/Software.html}.

\begin{appendix}
\section{Calculation of \texorpdfstring{$\mathbf{\lowercase{a}}'\bgamma+\bgamma'\mathbf{B}\bgamma$}{$a'gamma+gamma'Bgamma$}}\label{appA}
\begin{longlist}[2.]
\item[1. \textit{Two-dimensional square}.] For $V^2 (V>1)$ voxels on a
square, the $(V-2)^2$ voxels in the center all have 8 neighbors, the 4
vertex voxels have 3 neighbors, and the $4\cdot(V-2)$ voxels on the
edge but not vertexes have 5 neighbors. Then, given $\mathbf{a}$ and
$\mathbf{B}$ as defined in Section~\ref{secmodel}, we have
\begin{eqnarray*}
\mathbf{a}'\bgamma+\bgamma'\mathbf{B} \bgamma &=& a\cdot
V^2+b\cdot \bigl(8\cdot(V-2)^2+4\cdot3+5\cdot4\cdot(V-2)
\bigr)
\\
&=& (a+8b)V^2-12bV+4b.
\end{eqnarray*}

\item[2. \textit{Three-dimensional cube}.] For $V^3$ $(V>1)$ voxels in a cube,
the $(V-2)^3$ voxels in the center all have 26 neighbors, the 8 voxels
on the vertex have 7 neighbors, the $12(V-2)$ voxels on the edge but
not vertexes have 11 neighbors, and the $6(V-2)^2$ voxels on the 6
outside faces of the cube but not on the edges have 17 voxels. Then,
given $\mathbf{a}$ and $\mathbf{B}$ as defined in Section~\ref{secmodel}, we have
\begin{eqnarray*}
&& \mathbf{a}'\bgamma+\bgamma'\mathbf{B} \bgamma
\\
&&\qquad =a\cdot
V^3 +b\cdot \bigl(26(V-2)^3+8\cdot7+12(V-2)
\cdot11+6(V-2)^2\cdot17\bigr)
\\
&&\qquad = (a+26b) (V-2)^3+6(a+17b) (V-2)^2
+12(a+11b) (V-2)
\\
&&\qquad{} +8a+56b.
\end{eqnarray*}
\end{longlist}

\section{Posterior distributions in the Gibbs sampler}\label{appB}
\begin{longlist}[3.]
\item[1. \emph{Update} $\bgamma$.] We\vspace*{1pt} update the indicator for one voxel
$\gamma_j$ at a time. Let $\bgamma_{(-j)}= \{\gamma_l\dvtx  l \neq j\}$,
$I_{(-j)}$ be the set of indices $\{\gamma_l = 1\dvtx  l \neq j \} $,
$\bbeta_{(-j)}=\{\beta_l\dvtx  l\neq j\}$, and $\bX_{(-j)}$ be the design
matrix corresponding to $\bbeta_{(-j)}$. The prior probability of
$\gamma_j=1$, $\Pr(\gamma_j=1\mid \bgamma_{(-j)})$ is
$ \exp(a + b\sum_{l \in I_{(-j)}}{\gamma_l})/(1+\exp(a+ b\sum_{l
\in I_{(-j)}}{\gamma_l}))$.
By\vspace*{2pt} the Bayes rule, the posterior probability of $\gamma_j=1$ given the
data and other parameters is
\begin{eqnarray*}\label{eqprobgamma}
&& \Pr(\gamma_j=1\mid \bgamma_{(-j)},\bbeta, \sigma, \bY)
\\
&&\qquad = \frac{\Pr
(\gamma_j=1\mid \bgamma_{(-j)})}{\Pr(\gamma_j=1\mid \bgamma
_{(-j)})+F(j\mid \bgamma_{(-j)})^{-1} \cdot\Pr(\gamma_j=0\mid \bgamma
_{(-j)})},
\end{eqnarray*}
where $\bbeta\cdot\bgamma$ denotes the dot product between $\bbeta$
and $\bgamma$, and $F(j\mid \bgamma_{(-j)})$ is the Bayes factor,
\begin{eqnarray*}
F(j\mid \bgamma_{(-j)})&=&\frac{\Pr(\bY\mid \gamma_j=1,\bgamma_{(-j)},
\bbeta, \sigma)}{\Pr(\bY\mid \gamma_j=0,\bgamma_{(-j)}, \bbeta,
\sigma)}
\\
&=&\frac{\exp\{-\sum_{i=1}^n(Y_i-\bX_i\bbeta\cdot\bgamma
)^2/2\sigma^2\}}{\exp\{-\sum_{i=1}^n(Y_i-\bX_{i,(-j)}\bbeta
_{(-j)}\cdot\bgamma_{(-j)})^2/2\sigma^2\}},
\end{eqnarray*}
where $\bX_{i,(-j)}$ is the $i$th row of matrix $\bX_{(-j)}$.

\item[2. \emph{Update} $\sigma^2$.] $\sigma^2\mid - \sim\operatorname{Inv\mbox{-}Gamma}(n/2,\mu_{\sigma})$, where $\mu_{\sigma}=\sum_i({Y}_i-\mathbf{X}_i \bolds{\beta}\cdot\bolds{\gamma})^2/2$.

\item[3. \emph{Update} $\bbeta$.] Denote the $\beta_j$'s in $Z_j=h$ by
$\beta^h$, and let $\bX^h_i=\sum_{j\dvtx \gamma_j=1,Z_j=h} X_{ij}$. Note
that $\bX^h_i=0$ if $\{j\dvtx \gamma_j=1,Z_j=h\}=\varnothing$. Also, let
$\bbeta^{(-h)}=\{\beta_j\dvtx Z_j\neq h\}$, $\bolds{\gamma}^{(-h)}=\{\gamma
_j\dvtx Z_j\neq h\}$ and $\bX^{(-h)}=\{\bX_j\dvtx Z_j\neq h\}$, respectively,
denote the collection of all the $\beta$'s and the design matrix of the
covariates not in cluster $h$. Then for $h=1,\ldots, H$,
\[
\beta^h\mid - \sim\normal\bigl(\mu^h, 1/S^h
\bigr),
\]
with $S^h= \sum_{i=1}^n (\bX_i^h)^2/\sigma^2 +1/v^2$ and $\mu^h=\{
\sum_{i=1}^n (Y_i-\bX_i^{(-h)}\bbeta^{(-h)}\cdot\bolds{\gamma
}^{(-h)} )\bX_i^h\}/\break S^h$. This part can be parallelized (across $h$).

The posterior cluster membership $Z$ is drawn from a multinomial
distribution with
\begin{eqnarray*}
\Pr(Z_j = h\mid \gamma_j=1,-) &=& \frac{w_h \exp\{-\sum_{i=1}^n
(Y_i-\bX_i\bbeta_{(jh)}\cdot\bolds{\gamma}_{(jh)})^2/2\sigma^2\} }{
\sum_{k=1}^H w_k \exp\{-\sum_{i=1}^n (Y_i-X_i\bbeta_{(jk)}\cdot
\bolds{\gamma}_{(jk)})^2/2\sigma^2\}},
\\
\Pr(Z_j = h\mid \gamma_j=0,-) &=& w_h,
\end{eqnarray*}
where $\bbeta_{(jh)}=(\beta_1,\ldots,\beta_{j-1},\beta^{h},\beta
_{j+1}, \ldots,\beta_p)$ and $\bolds{\gamma}_{(jh)}=(\gamma_1,\ldots,\gamma
_{j-1},1, \gamma_{j+1},\break\ldots,\gamma_p)$ for $h=1,\ldots,H$ and $j=1,\ldots,
p$. To update the associated weights $\bw$, first set $w_H' = 1$ and
draw $w_h'$ from $\operatorname{Beta}  (1 + \sum_{j\dvtx  Z_j=h} 1, \alpha +
\sum_{j\dvtx  Z_j>h}
1 )$ for each $h \in\{1, \ldots, H-1\}$, then update $w_h =
w_h' \prod_{k<h}(1- w_k')$.
\end{longlist}
\end{appendix}

\section*{Acknowledgments}
The authors are grateful to the Associate Editor and four reviewers for
constructive comments that helped improve the exposition and clarity of
the paper, and to Nancy Zhang for insightful discussions. The content
is solely the responsibility of the authors and does not necessarily
represent the official views of NIMH, the National Institutes of Health
or SAMSI.

Part of the project was conducted when Fan Li and Tingting Zhang were
research fellows of the Object
Data Analysis program of the U.S. Statistical and Applied Mathematical Sciences
Institute (SAMSI).

\begin{supplement}[id=suppA]
\stitle{Heatmaps}
\slink[doi]{10.1214/15-AOAS818SUPP} 
\sdatatype{.pdf}
\sfilename{aoas818\_supp.pdf}
\sdescription{We provide the heat\-maps of the voxels with top 10\%
highest posterior selection probabilities obtained, resulting from
Ising-DP, Ising-Gaussian and i.i.d.-Gaussian priors, respectively, in three regressions [\citet{supp}].}
\end{supplement}

%

\printaddresses
\end{document}